\documentclass[aps,prb,groupedaddress,showpacs,twocolumn]{revtex4}
\usepackage{graphicx}
\usepackage{amsmath}
\usepackage{amssymb}
\usepackage{bbm}
\usepackage{xspace}
\usepackage{color}
\usepackage{hyperref}

\newcommand{\matx}{\mat{x}}
\newcommand{\GF}{\text{G}}
\newcommand{\gf}{\text{g}}
\newcommand{\TF}{\text{T}}

\definecolor{orange}{RGB}{252,77,6}
\definecolor{brown}{RGB}{200,127,50}
\definecolor{green1}{RGB}{00,100,00}
\definecolor{green2}{RGB}{00,150,00}
\definecolor{green3}{RGB}{00,200,00}
\definecolor{green4}{RGB}{00,250,00}
\definecolor{blue1}{RGB}{102,204,255}
\definecolor{sth}{RGB}{50,200,160}

\newcommand{\fig}[1]{fig.\thinspace{}\ref{#1}}

\newcommand{\eq}[1]{eq.\thinspace{}(\ref{#1})}
\newcommand{\Eq}[1]{Eq.\thinspace{}(\ref{#1})}

\newcommand{\se}{sec.\@\xspace}

\newcommand{\app}{app.\@\xspace}

\newcommand{\etal}[0]{\textit{et al.}}
\newcommand{\tcite}[1]{ref.~\onlinecite{#1}}
\newcommand{\tcites}[1]{refs.~\onlinecite{#1}}

\newcommand{\tr}[1]
{
\text{tr}\,#1 
}

\newcommand{\uu}{1\hspace{-3pt}1}

\DeclareMathOperator{\sign}{sign}

\newcommand{\nag}{{\phantom{\dag}}}

\newcommand{\ve}[1]{{\bf #1}}
\newcommand{\mat}[1]{\mathsf{#1}}

\begin{document}


\title{Steady-state spectra, current and stability diagram of a quantum dot:\\ a non-equilibrium Variational Cluster Approach}


\author{Martin Nuss}
\email[]{martin.nuss@student.tugraz.at}
\affiliation{Institute of Theoretical and Computational Physics, Graz University of Technology, 8010 Graz, Austria}
\author{Christoph Heil}
\affiliation{Institute of Theoretical and Computational Physics, Graz University of Technology, 8010 Graz, Austria}
\author{Martin Ganahl}
\affiliation{Institute of Theoretical and Computational Physics, Graz University of Technology, 8010 Graz, Austria}
\author{Michael Knap}
\affiliation{Institute of Theoretical and Computational Physics, Graz University of Technology, 8010 Graz, Austria}
\author{Hans Gerd Evertz}
\affiliation{Institute of Theoretical and Computational Physics, Graz University of Technology, 8010 Graz, Austria}
\author{Enrico Arrigoni}
\affiliation{Institute of Theoretical and Computational Physics, Graz University of Technology, 8010 Graz, Austria}
\author{Wolfgang von der Linden}
\affiliation{Institute of Theoretical and Computational Physics, Graz University of Technology, 8010 Graz, Austria}


\date{\today}

\begin{abstract}
We calculate steady-state properties of a strongly correlated quantum dot under voltage bias by means of non-equilibrium Cluster Perturbation Theory and the non-equilibrium Variational Cluster Approach, respectively. Results for the steady-state current are benchmarked against data from accurate Matrix Product State based time evolution. We show that for low to medium interaction strength, non-equilibrium Cluster Perturbation Theory already yields good results, while for higher interaction strength the self-consistent feedback of the non-equilibrium Variational Cluster Approach significantly enhances the accuracy. We report the current-voltage characteristics for different interaction strengths. Furthermore we investigate the non-equilibrium local density of states of the quantum dot and illustrate that within the variational approach a linear splitting and broadening of the Kondo resonance is predicted which depends on interaction strength. Calculations with applied gate voltage, away from particle hole symmetry, reveal that the maximum current is reached at the crossover from the Kondo regime to the doubly-occupied or empty quantum dot. Obtained stability diagrams compare very well to recent experimental data [\href{http://link.aps.org/doi/10.1103/PhysRevB.84.245316}{Phys. Rev. B,{\bf 84}, 245316 (2011)}].
\end{abstract}


\maketitle

\section{Introduction}\label{sec:introduction}
The understanding of non-equilibrium phenomena in strongly correlated many-body systems may reveal previously unknown fundamental aspects of physics as well as prove crucial for the development of technical applications in the fields of nano or molecular electronics. Currently, combined insight of experiments on nano-devices~\cite{goldhaber_from_1998, PhysRevB.84.245316, Kretinin_2012} and results from artificial quantum simulators (see e.g. \tcites{Brantut2012, schneider_fermionic_2012, Trotzky18012008, PhysRevLett.104.080401, greiner_collapse_2002}) are capable of providing coherent insight into the non-equilibrium behavior of the quantum world. These experiments provide both a challenge for theoretical concepts as well as an accurate check for theoretical results. Both nano-devices and condensed matter simulators are often described remarkably well by interacting model Hamiltonians which are in general not exactly solvable.\\
Here we focus on a model of a single quantum dot, the single-impurity Anderson model (SIAM)~\cite{anderson_localized_1961}. This model, incorporating spin and charge fluctuations as well as Kondo correlations, has been studied as an idealized realization of an interacting system by a wide array of techniques in equilibrium (for an overview see e.g. Hewson~\tcite{hewson_kondo_1997}). The model we use here is applicable to generic single quantum dot systems including effects of a finite lead bandwidth.\\
However, the evaluation of dynamic quantities of strongly correlated quantum many-body systems out of equilibrium poses a notoriously difficult problem. A particular challenge for the SIAM is that it is expected to remain in a strong coupling regime, even under the influence of a bias voltage~\cite{coleman_is_2001}. Additional challenges are posed by the need for particle and/or energy dissipation mechanisms. Techniques which allow the calculation of physical quantities beyond the linear response regime are quite restrictive up to now and no full understanding of the non-equilibrium dynamics or the steady-state under bias are available for the SIAM, apart from some special cases~\cite{PhysRevLett.80.4370,PhysRevB.62.R16271,PhysRevB.79.245102}. Nevertheless, controlled results have been obtained with the analytical Bethe Ansatz for some physical quantities in the interacting resonant level model~\cite{mehta_nonequilibrium_2006} and for the SIAM~\cite{PhysRevLett.87.236801}. Perturbative calculations~\cite{PhysRevB.68.155310, PhysRevLett.67.3720} as well as non-crossing approximation studies~\cite{PhysRevLett.70.2601, PhysRevB.49.11040} extend early fundamental work on the non-equilibrium problem~\cite{meir_transport_1991, meir_landauer_1992, meir_low-temperature_1993, jauho_time-dependent_1994} and the impurity out of equilibrium~\cite{ng_onsite_1988, aguado_ac_1998, kaminski_universality_2000}. Depending on the setup of parameters and model details, insight may be gained by semi-classical methods~\cite{richter_semiclassical_1999} or master equation approaches~\cite{timm_tunneling_2008}. Recently moreover, several techniques, which have proven very successful in the equilibrium theory, have been extended to the non-equilibrium case. Among them are many-body cluster methods~\cite{balzer_non-equilibrium_2011,knap_nonequilibrium_2011}, renormalization group (RG) approaches~\cite{1367-2630-10-4-045012,jakobs_nonequilibrium_2010,PhysRevB.73.245326, PhysRevLett.84.3686, PhysRevLett.87.156802, PhysRevB.75.045324, PhysRevLett.99.150603}, flow equation methods~\cite{wang_flow_2010, PhysRevLett.95.056602}, real time path integral calculations~\cite{PhysRevB.77.195316}, out of equilibrium noncrossing approximation (NCA) ~\cite{schiller_measuring_2001}, generalized slave-boson methods~\cite{PhysRevLett.85.1946, PhysRevB.84.125303}, diagrammatic quantum Monte Carlo (QMC)~\cite{PhysRevB.81.035108,  PhysRevB.79.035320, PhysRevB.79.153302} or QMC methods based on a complex chemical potential~\cite{PhysRevLett.99.236808, Dirks2011, Dirks2012a, Dirks2012b}. The Gutzwiller approximation has been generalized to the time-dependent case~\cite{Fabrizio2012} and so has numerical renormalization group (NRG)~\cite{PhysRevLett.101.066804, PhysRevLett.95.196801, PhysRevB.74.245113, PhysRevLett.100.087201} where however some issues with the use of Wilson chains in non-equilibrium systems have been pointed out by Rosch~\cite{rosch_wilson_2011}. Dual fermion approaches~\cite{jung_dualfermion2010} have been proposed as well as super operator techniques~\cite{Dutt2011, mu_universal_2011}. Some recent work attempts to compare several of these theories~\cite{0957-4484-21-27-272001, 1367-2630-12-4-043042, uimonen_comparative_2011} and shed light on the critical issue of time scales involved~\cite{PhysRevB.85.075301}. Finally, some results for the SIAM are available~\cite{heidrich_nonequilibrium_2010} from numerically exact time evolution by a combination of Density Matrix Renormalization Group (DMRG)~\cite{PhysRevB.48.10345} and successive time evolution via time-dependent DMRG (tDMRG)~\cite{heidrich_nonequilibrium_2010, PhysRevB.73.195304, PhysRevB.78.195317, Schneider2006}. They are currently limited to small bias voltages and moderate interaction strengths.\\
In the present paper we explore the capabilities of non-equilibrium Cluster Perturbation Theory (nCPT)~\cite{balzer_non-equilibrium_2011} and benchmark the non-equilibrium Variational Cluster Approach (nVCA)~\cite{knap_nonequilibrium_2011} on the SIAM. We obtain the full current-voltage characteristics which we compare to results from a very accurate time evolution by means of Time Evolving Block Decimation (TEBD)~\cite{unpubTEBD} and QMC~\cite{PhysRevB.81.035108}. We also comment on the comparison to other results obtained by tDMRG~\cite{PhysRevB.79.235336}, perturbation theory~\cite{PhysRevB.68.155310} and FRG~\cite{jakobs_nonequilibrium_2010}. Because of good agreement with these data, we then proceed to evaluate the single-particle spectrum of the quantum dot in the steady-state. Since this is a dynamic quantity, it is even harder to obtain for most numerical methods. We find a linear splitting of the Kondo resonance~\cite{PhysRevLett.89.156801, PhysRevLett.95.126603} which depends on the interaction strength. Detailed results for the particle-hole symmetric model where Kondo correlations dominate are presented and supplemented by data in the whole parameter space including an applied gate voltage. We highlight the crucial edge which nVCA gains over nCPT via its self-consistent feedback.\\
Both many-body cluster techniques, nCPT and nVCA are based on the well established equilibrium counterparts Cluster Perturbation Theory (CPT)~\cite{gros_cluster_1993, senechal_spectral_2000} and the Variational Cluster Approach (VCA)~\cite{potthoff_variational_2003,Potthoff2011}. They make use of the non-equilibrium Keldysh-Schwinger Green's function technique~\cite{Keldysh1965, feynman_theory_1963, schwinger_brownian_1961}. The present paper extends and benchmarks the ideas for adapting VCA to non-equilibrium situations introduced in \tcite{knap_nonequilibrium_2011}. We attempt to give a comprehensive overview of the current capabilities and shortcomings of nCPT and nVCA for the application to steady-state problems of strongly correlated systems. The SIAM provides an excellent probing ground for our purposes as a model where the effects of correlations are crucial. It is however still relatively simple which permits systematic study and some results are available which allow for comparison. Reasonable results for the SIAM in equilibrium have been obtained previously by CPT as well as VCA~\cite{PhysRevB.85.235107}. Both cluster methods are approximate but they yield fairly reliable results and are therefore interesting due to their great flexibility and versatility. They are computationally not very demanding and allow in principle to treat a wide range of fermionic / bosonic lattice Hamiltonians out of equilibrium. Possible extensions include electronic multi-band or multi-orbital systems in one, two or three dimensions also including phonons.\\
This paper is organized as follows: In \se~\ref{sec:model} we sketch the SIAM and describe the setup used. We proceed by outlining nCPT (\se~\ref{ssec:methodsNCPT}) and nVCA (\se~\ref{ssec:methodsNVCA}). In \se~\ref{ssec:resultsSSCurrent}, a comparison of steady-state currents as obtained by nCPT, nVCA, DMRG and successive TEBD~\cite{PhysRevLett.93.040502} and QMC~\cite{PhysRevB.81.035108} is presented. The non-equilibrium local density of states (nLDOS) is examined in \se~\ref{ssec:nLDOS}. Finally effects of an applied gate voltage on the steady-state current are studied in \se~\ref{ssec:resultsGatedCurrent}.\\

\section{Model of a single quantum dot}\label{sec:model}
\begin{figure}
	\centering
		\includegraphics[width=0.48\textwidth]{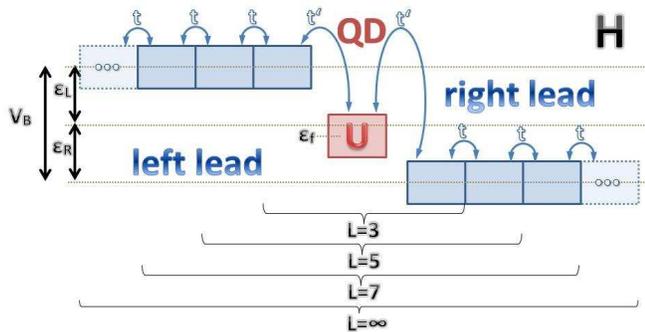}
	\caption{(Color online) Illustration of the SIAM as prepared for use within nCPT and nVCA. An interacting quantum dot (QD) is located in the middle of two non-interacting electronic leads. In order to employ the cluster approaches, the system is divided into three pieces, a left part, a right part and a central region of length $L$, which includes the quantum dot as well as parts of left and right leads.}
	\label{fig:Model}
\end{figure}

We model the setup, consisting of a single correlated quantum dot in-between two metallic leads, by a single-site Hubbard model embedded in a one dimensional infinite tight-binding chain. The Hamiltonian of this single-impurity Anderson model (SIAM)~\cite{anderson_localized_1961} reads (see \fig{fig:Model})
\begin{subequations}
\begin{align}
\hat{\mathcal{H}} &= \hat{\mathcal{H}}_{\text{dot}} + \hat{\mathcal{H}}_{\text{lead}} + \hat{\mathcal{H}}_{\text{coup}}\label{eq:HSIAM}\\
\hat{\mathcal{H}}_{\text{dot}} &= \epsilon_f \, \sum\limits_{\sigma}  \, f_{\sigma}^\dagger \, f_{\sigma}^\nag + U \, \hat{n}^{f}_{\uparrow} \, \hat{n}^{f}_{\downarrow}\label{eq:Hdot} \\
\hat{\mathcal{H}}_{\text{lead}} &= \sum\limits_{\alpha,\sigma}\,\left( \epsilon_{\alpha} \, \sum\limits_{i=0}^{\infty}  \,c_{i\alpha\sigma}^{\dagger} \, c_{i\alpha\sigma}^{\nag} - t \, \sum\limits_{\left\langle i,\,j \right\rangle}  \, c_{i\alpha\sigma}^{\dagger} \, c_{j\alpha\sigma}^{\nag}\right)\label{eq:Hlead}\\
\hat{\mathcal{H}}_{\text{coup}} &= -t'\, \sum\limits_{\alpha,\sigma} \left( c_{0\alpha\sigma}^{\dagger} \, f_{\sigma}^\nag + f_{\sigma}^\dagger \, c_{0\alpha\sigma}^{\nag} \right)\label{eq:Hcoupl}\;\mbox{.}
\end{align}
\end{subequations}
The electronic annihilation (creation) operators $c_{i\alpha\sigma}^{\nag}, f_{\sigma}^{\nag}$ ($c_{i\alpha\sigma}^{\dag}, f_{\sigma}^{\dag}$) obey the usual anti-commutation relations with spin $\sigma = \{\uparrow, \downarrow\}$ and annihilate (create) electrons in the left or right lead $\alpha=\{L,R\}$ or the quantum dot, respectively. The particle number operator of the quantum dot is defined by $\hat{n}^{f}_{\sigma}=f_\sigma^\dag f_\sigma^\nag$. $U$ represents the on-site Hubbard repulsion. The single-particle energy of the quantum dot $\epsilon_f = -\frac{U}{2} + V_G$ is comprised of a gate voltage $V_G$ and a shift by $-\frac{U}{2}$ (\eq{eq:Hdot}), such that the particle-hole symmetric point is reached when the gate voltage vanishes and the bias is applied in an anti-symmetric fashion as described below. Non-interacting left and right leads are described by tight binding chains with a nearest-neighbor hopping $t$ (\eq{eq:Hlead}). The lead on-site energies are denoted by $\epsilon_{L/R}$. A bias voltage $V_B$ may be applied to the system in an anti-symmetric fashion by setting $\epsilon_L = \mu_L = -\epsilon_R = -\mu_R = \frac{V_{B}}{2}$, where $\mu_{L/R}$ denote the chemical potentials of the decoupled leads ($t'=0$). Finally, the symmetric tunneling between the non-interacting leads and the quantum dot is denoted by $t'$ (\eq{eq:Hcoupl}). In this work we adopt units in which $\hbar, e$, as well as the inter-lead hopping $t$ are equal to $1$.  The lead-dot tunneling is fixed to $t'=0.3162$ throughout this paper, which implies an Anderson width of $\Delta \equiv \pi\,t'^2\,\rho_{\text{lead}}(0) = \frac{t'^2}{t} \approx 0.1$. This choice of parameters furthermore implies an effective lead bandwidth of $D=40\,\Delta$. For higher bias voltages we therefore include additional effects due to finite bandwidth and band-shape as compared to the often used wide band limit. The prefactor for the current is then given by $j_0=\frac{2|e|t}{\hbar}$. In the following, we always consider the zero temperature case.\\

\section{Methods}\label{sec:methods}
All information needed to evaluate expectation values in the steady-state of an interacting fermionic lattice Hamiltonian is contained in a single object, the single-particle Green's function in Keldysh space $\widetilde{\GF}$ (\eq{eq:Gtilde}) (for details see \app~\ref{app:basics}). Evaluating this object is in general impossible to do exactly but a handle on $\widetilde{\GF}$ is given by nCPT and nVCA. These approximations are to be discussed in the following.\\

\subsection{Non-equilibrium cluster perturbation theory}\label{ssec:methodsNCPT} 
The idea of nCPT is to split an (infinitely) large system $\hat{\mathcal{H}}$ for times $\tau<\tau_0$ into a set of small decoupled clusters (described by the Hamiltonian $\hat{h}$), for which the single-particle Green's functions can be determined exactly, by numerical means
\begin{align*}
\hat{\mathcal{H}} = \hat{h} + \theta(\tau - \tau_0)\, \hat{\TF}\,\mbox{.} 
\end{align*}
At time $\tau_0$ the coupling of the individual subsystems $\hat{\TF} = \hat{\mathcal{H}}-\hat{h}$ is switched on and the solution for the single-particle Green's function of the total system can be obtained by the CPT equation~\cite{gros_cluster_1993} in Keldysh-, site- and spin-space
\begin{align}
 \widetilde{\GF}^{-1} &= \widetilde{\gf}^{-1}-\hat{\TF} \otimes \tilde{\uu}\,\mbox{,}
\label{eq:CPT}
\end{align}
where $\widetilde{\GF}/\widetilde{\gf}$ denote the single-particle Green's function in Keldysh space of the total system/split systems. The unit matrix in Keldysh space is denoted by $\tilde{\uu}$. \Eq{eq:CPT} is a strong coupling perturbation theory result and holds up to first order in the inter-cluster hopping $\hat{\TF}$, as far as the self-energy is concerned. As a consequence, the self-energy $\Sigma_{\hat{\mathcal{H}}}$ of the total system is approximated by the self-energy of the initial system $\Sigma_{\hat{h}}$. This implies that nCPT/nVCA become exact in the non-interacting limit, since in this case, the self-energy of the total system agrees with that of the initial system, which are both zero. The results of nCPT and nVCA converge towards the exact results with increasing cluster size.\\
In the present paper we illustrate this approach for the SIAM out of equilibrium. We start out by splitting the infinite chain (\eq{eq:HSIAM}) into three parts: i) an interacting central region of length $L$ consisting of the interacting site (the quantum dot) as well as an equal amount of sites of the left and the right leads, ii) a semi-infinite, non-interacting left region consisting of the remaining part of the left lead and iii) a semi-infinite, non-interacting right region consisting of the remaining part of the right lead (see \fig{fig:Model}). To proceed, the single-particle Green's functions in Keldysh space have to be determined exactly for those three systems. Results for the left and right part are available analytically by the retarded Green's function of a semi-infinite tight binding chain~\cite{economou_greens_2010}
\begin{align}
 \gf^{L/R}_{i,j}(\omega) &= \upsilon^{L/R}_{0,i-j}(\omega) - \upsilon^{L/R}_{0,i+j}(\omega)   \label{eq:env}\\
  \upsilon^{L/R}_{i,j}(\omega) &= \frac{-\imath}{\sqrt{4 |t|^2-(\omega-\epsilon^{L/R})^2}}\,\Bigg( -\frac{\omega-\epsilon^{L/R}}{2|t|} \label{eq:envGF}\\
\nonumber  &+ \imath\,\sqrt{1-\left( \frac{\omega-\epsilon^{L/R}}{2|t|} \right)^2}\Bigg)^{|i-j|}\;\mbox{,}
\end{align}
where $\upsilon_{i,j}$ is the retarded Green's function of the infinite tight binding chain. The single-particle Green's function of the central interacting region can be determined in the Q-matrix formalism by the Band-Lanczos algorithm (see \tcite{PhysRevB.85.235107} and \tcite{PhysRevB.65.045109}). The advanced component can always be obtained by the relation $\GF^A=(\GF^R)^{\dag}$. Before coupling the three subsystems at time $\tau=\tau_0$, each of them is in equilibrium at different chemical potentials $\mu_{L/R/C}$. Therefore we may evaluate the corresponding Keldysh components by the relation~\cite{negele_quantum_1998}
\begin{align}
 \GF^K(\omega,\mu) &= \left(\GF^R(\omega)-\GF^A(\omega)\right)\left(1-2\,p_{\text{FD}}(\omega,\mu)\right)\,\mbox{,}
\label{eq:GFK}
\end{align}
where $p_{\text{FD}}$ denotes the Fermi-Dirac distribution function at inverse temperature $\beta$: $p_{\text{FD}} = \frac{1}{1+e^{\beta(\omega-\mu)}}$. In the zero temperature limit, $\left(1-2\,p_{\text{FD}}(\omega,\mu)\right)$ may be re-expressed as $\sign(\omega-\mu)$. This is the only expression in which the chemical potential $\mu$ enters. It is crucial that this relation does not hold in a non-equilibrium situation any longer. The hermitian part of $\GF^K(\omega,\mu)$ is zero in equilibrium. The imaginary part consists of contributions due to delta peaks for finite size systems like the central regions. The operator $\hat{\TF}$ just contains the two hopping terms from the left to the central and from the central to the right region. At time $\tau_0$, the hopping processes between these three regions are switched on and the steady-state single-particle Green's function $\widetilde{\GF}$ is determined using nCPT, \eq{eq:CPT}. As in the equilibrium case, the CPT results can be improved by the variational cluster approach, in which the single-particle part of the initial system is suitably modified. In the following a variational extension (nVCA) of the scheme described above will be presented following \tcite{knap_nonequilibrium_2011}.\\

\subsection{Non-equilibrium variational cluster approach}\label{ssec:methodsNVCA} 
The idea of enlarging the size of the central region in nCPT is to find a better approximation for the starting point of perturbation theory. As pointed out before, in CPT the self-energy of the total system is approximated by that of the decoupled system. By increasing the size of the central region, the self-energy of the decoupled system converges gradually towards that of the total system. In nVCA an even more suitable initial system is chosen. This can be achieved by making use of the fact that the decomposition of $\hat{\mathcal{H}}$ into clusters $\hat{h}$ and inter-cluster parts $\hat{\TF}$ is not unique. We are free to add single-particle operators $\hat{\Delta}(\matx)$, which depend linearly on parameters $\matx$, to the initial hamiltonian $\hat{h}$, provided we subtract them again from $\hat{\TF}$
\begin{align*}
 \hat{\mathcal{H}} &= \left(\hat{h} + \hat{\Delta}(\matx)\right) + \theta(\tau - \tau_0)\,\left(\hat{\TF} - \hat{\Delta}(\matx)\right)\,\mbox{.}
\end{align*}
Here,
\begin{align*}
 \hat{\Delta}(\matx) &= \sum\limits_l\,\matx_l\,\hat{\Delta}_l\,\mbox{,}
\end{align*}
where $\hat{\Delta}_l$ is quadratic in the fermion operators. This parametrized single-particle field introduces additional freedom in the starting guess for perturbation theory and can be used for a self-consistent feedback on the clusters. The system described by $\left(\hat{h} + \hat{\Delta}(\matx)\right)$, usually referred to as reference system in the context of VCA, has the same structure as in equilibrium VCA. The condition to fix the single-particle parameters $\matx$, however, is different in nVCA and VCA. Equilibrium VCA is based on the Self-energy Functional Approach (SFA)~\cite{potthoff_self-energy-functional_2003, potthoff_self-energy-functional_2003-1} and provides a variational principle for the generalized grand potential functional, which is not well defined in the context of non-equilibrium systems any longer. An alternative criterion for fixing the variational parameters $\matx$ was introduced in \tcite{knap_nonequilibrium_2011} and compared to the traditional VCA criterion in \tcite{PhysRevB.85.235107}. It is based on the idea of starting from a system which is as similar as possible to the total, original system in terms of physically observable quantities. We demand the expectation values of the operators $\hat{\Delta}_l$ to coincide in the initial (reference) system and the steady-state of the total system, i.e. 
\begin{align*}
\langle \hat{\Delta}_l\rangle_{\gf}&\stackrel{!}{=}\langle \hat{\Delta_l}\rangle_{\GF}\,\mbox{.}
\end{align*}
For example, adding variational freedom in the on-site energy of the quantum dot, corresponding to  $\Delta = \matx_{\epsilon_f}\, \hat{n}_\sigma^f$ yields the self-consistency condition: $\langle \hat{n}_\sigma^f\rangle_{\gf}\stackrel{!}{=}\langle \hat{n}_\sigma^f\rangle_{\GF}$. From a more conceptual point of view, it is interesting that these self-consistency conditions follow from the condition~\cite{knap_nonequilibrium_2011}
\begin{align}
\int_{-\infty}^\infty \frac{d\omega}{2\pi}\tr\left\{  
\widetilde{\tau}_1 \frac{\partial \left(\widetilde{\gf}_{0}\right)^{-1}}{\partial \ve \matx'} 
\left(\widetilde{\gf}-\widetilde{\GF}\right)\right\}
= 0 \,\mbox{,}
\label{eq:SC}
\end{align}
which is closely related to the stationarity condition of the generalized grand potential functional $\Omega$ in the equilibrium approach~\cite{sene_2009}. Here $\widetilde{\tau}_1= \begin{pmatrix} 0 \uu\\ \uu 0 \end{pmatrix} $ is a Pauli matrix in Keldysh space and the subscript zero denotes non-interacting Green's functions. From the numerical point of view, one has to find the roots of an n-dimensional set of non-linear equations (where n is the number of variational parameters $\matx$).\\
Like in first order Dirac perturbation theory, the influence of the perturbation increases with time (perturbations introduced due to the coupling with hopping elements $t$ at $\tau_0$ are proportional to $t \cdot (\tau-\tau_0)$ since we are considering first order perturbation theory) and one might argue that nCPT is bound to fail in the long-time, steady-state limit, even for small couplings between the clusters~\cite{balzer_non-equilibrium_2011}. This argument cannot be true in general, as nCPT yields exact results in the case of non-interacting particles, although the initial decoupled systems are far from the steady-state behavior.\\
The initial system in nCPT is independent of the non-equilibrium situation and this shortcoming is improved to some extent within nVCA, where the information about the applied bias voltage is self-consistently fed back to the initial reference system. As we will see in this paper, there are circumstances under which nCPT already yields reasonable results. In general, however, we observe that nVCA represents a significant improvement over nCPT. This implies, on the one hand, that under steady-state conditions, the self-energy in the central cluster is significantly modified when going from nCPT to nVCA. On the other hand, it indicates that the steady-state situation can be mimicked in an equilibrium system by the auxiliary one-particle terms, determined self-consistently. A point for improvement of this approach is to modify the self-energy so that it is a genuine non-equilibrium one.\\
Details on the nVCA procedure and the particular choice of variational parameters are provided in \app~\ref{app:nvca}.\\
\begin{figure*}
	\centering
		\includegraphics[width=1.00\textwidth]{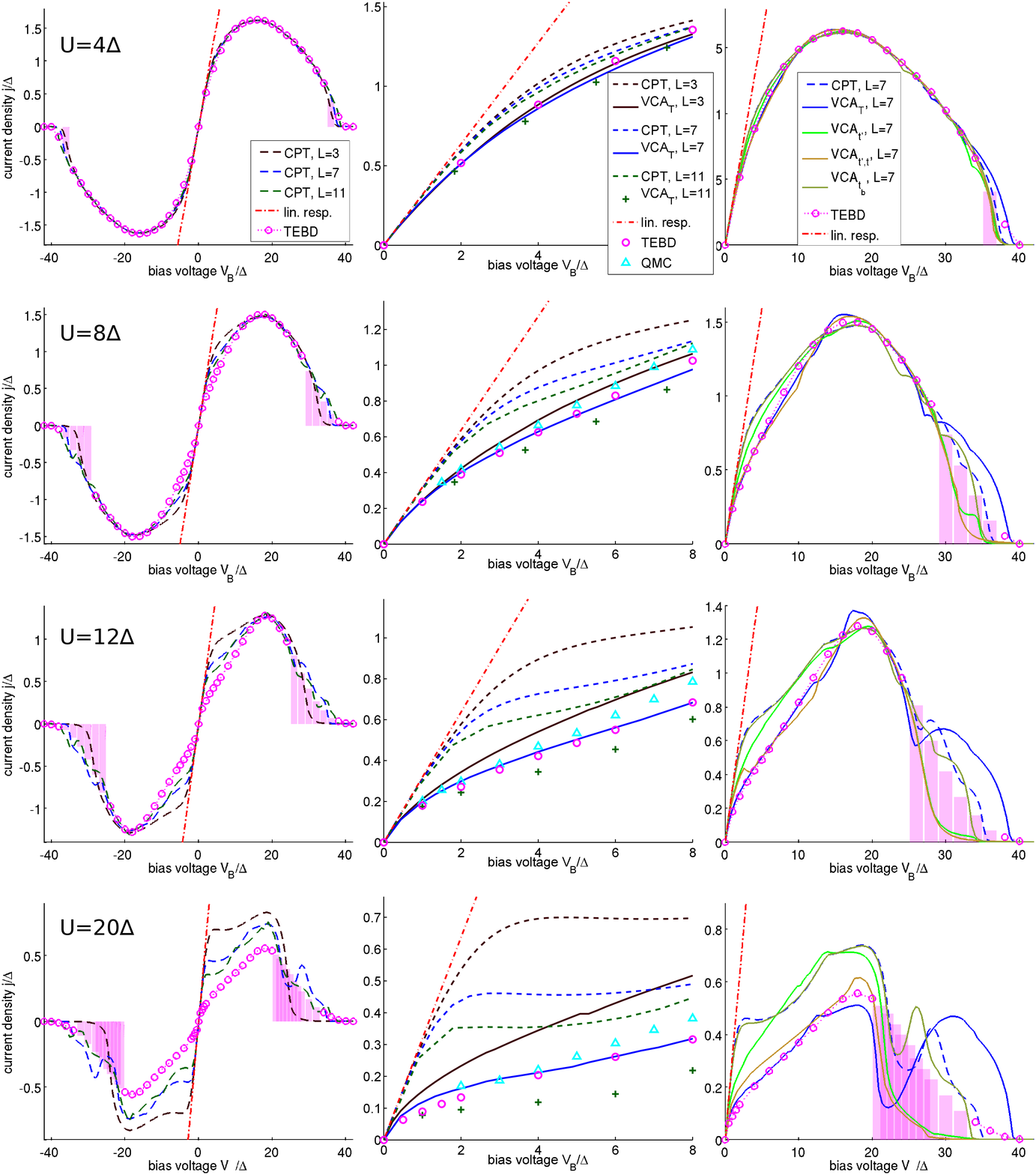}
	\caption{(Color online) Steady-state current-voltage characteristics in the particle-hole symmetric case ($V_G=0$). Results are shown for small to large interaction strength: $U=4\,\Delta$ (top row), $U=8\,\Delta$ (second row), $U=12\,\Delta$ (third row) and $U=20\,\Delta$ (bottom row). The first column contains the full current-voltage characteristics as obtained with nCPT for $L=3,7 \mbox{ and } 11$. As benchmarks, accurate TEBD results~\cite{unpubTEBD} and the linear response result $j_{\text{lin}}=2\,G_0\,V_{B}$ are depicted as well. In some bias regions, only an upper bound can be obtained by TEBD. This upper limit is depicted as a magenta bar. In the second column a close-up of the low bias region is presented. Here we compare nCPT ($L=3, 7 \mbox{ and } 11$), nVCA$_T$ ($L=3, 7 \mbox{ and } 11$) with TEBD data as well as QMC results~\cite{PhysRevB.81.035108}. Note that the QMC data have been obtained in a wide-band limit. The third column contains results obtained for $L=7$ for various sets of nVCA variational parameters: nCPT, nVCA$_T$, nVCA$_{t'}$, nVCA$_{t,t'}$, nVCA$_{t_b}$. Legends are displayed once in the top row for the respective column.}
	\label{fig:ssCurrent}
\end{figure*}

\section{Results}\label{sec:results}
In the following we compare nCPT and nVCA data for the steady-state current with TEBD~\cite{unpubTEBD} and QMC~\cite{PhysRevB.81.035108} results in the particle-hole symmetric model. We elaborate on the non-equilibrium density of states and finally discuss results for the steady-state current away from particle-hole symmetry by applying a gate voltage. Earlier preliminary results obtained for the SIAM out of equilibrium by nCPT and nVCA are available in \tcite{Nuss2011} and \tcite{Nuss2012}.\\

\subsection{Steady-state current}\label{ssec:resultsSSCurrent}
Here we investigate the steady-state current-voltage characteristics of the particle-hole symmetric SIAM. In this parameter region, Kondo correlations are especially important. All nCPT and nVCA results are for an infinite system using a self-energy based on an $L\leq11$ site interacting reference system. Although the length of Kondo correlations scales exponentially in interaction strength, it has been shown in \tcite{PhysRevB.85.235107} that the VCA approximation, similarly to other approximate methods such as FRG or Gutzwiller wave functions, is capable of retaining qualitatively most features of the ongoing Kondo physics, although, possibly, with renormalized scales. The steady-state current can be evaluated on any link either within the central region or between the central and the neighboring regions, since we find that the continuity equation is fulfilled within at least $10^{-6}$ relative to the steady-state current amplitude. We note again that in the non-interacting case nCPT as well as nVCA become exact so we do not show these data explicitly.\\
\begin{figure*}
	\centering
		\includegraphics[width=1.00\textwidth]{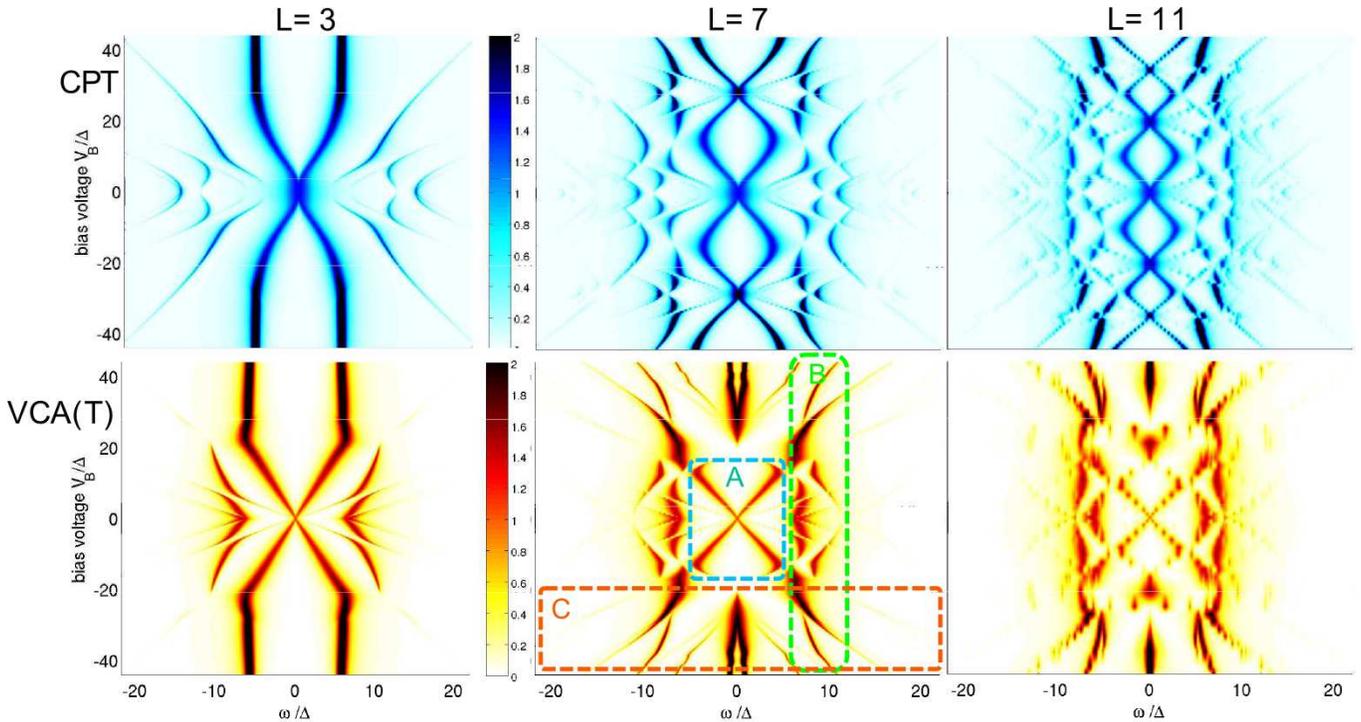}
	\caption{(Color online) Non-equilibrium local density of states of the quantum dot $\rho_f$ as a function of energy $\omega$ and bias voltage $V_B$ in the particle-hole symmetric case ($V_G=0$). Region A represents the experimentally most interesting region where no effects of the lead bands come into play and one observes a splitting of the equilibrium Kondo resonance. Region B is dominated by high energy incoherent excitations which are expected to form continuous Hubbard bands in the thermodynamic limit. In region C, the lead density of states becomes relevant, causing a new low energy resonance for $L>=7$. Data are depicted for $U=12\,\Delta$ and an artificial numerical broadening of $0^+=0.05$. nCPT results are shown in the first row for system sizes of $L=3$ (left), $L=7$ (mid) and $L=11$ (right). The second row displays the corresponding nVCA$_{T}$ data. For reasons of visualization the color-scale is cutoff at $\rho_{\text{max}} = 2$. Note that a horizontal line through the center of each plot at $V=0$ corresponds to the equilibrium density of states. The spectral function sum rule \eq{eq:ASR} is fulfilled for  each applied bias voltage. The $L=11$ results are noisy as fewer data points have been obtained.}
	\label{fig:NDOS}
\end{figure*}
Results obtained by \eq{eq:sscurrent} for the nCPT case are shown in the left column of \fig{fig:ssCurrent} for various values of the interaction strength $U$. The linear response result $j_{\text{lin}}=2\,G_0\,V_{B}$, where $G_0$ is the conductance quantum, is fulfilled within nCPT (and nVCA) in the very low bias region. This is because these methods fulfill the Friedel sum rule in equilibrium~\cite{PhysRevB.85.235107}. Actually, nCPT tends to stay in the linear response regime longer than nVCA and therefore overestimates the current for low (but not very low) bias voltages. We model the leads by one dimensional tight binding chains, yielding a semi-circular density of states (\eq{eq:envGF}) of bandwidth $D=40\,\Delta$. This implies a vanishing current in the high bias limit at $V_{B}=40\,\Delta$, due to non-overlapping lead density of states. This limit again is trivially fulfilled within nCPT and nVCA. Note that in a wide band limit the current curves would approach a constant roughly at their respective maxima in the data shown.\\
For comparison, numerically exact results~\cite{unpubTEBD} as obtained by a real-time evolution with TEBD are also indicated in \fig{fig:ssCurrent}. Note that the TEBD data are obtained from the steady-state plateau in the time dependence of the current. At small and medium bias, the current converges well and the TEBD results provide an accurate benchmark. At higher bias, convergence is low, but the TEBD data still provide a reasonably reliable upper bound for the steady-state current.\\
In the intermediate bias region it is interesting to investigate the behavior of nCPT with increasing size of the central region $L$. As can be seen from the plots, for any interaction strength increasing $L$ yields monotonically improving results. While for low interaction strength $U=4\,\Delta$ the nCPT results almost coincide with the TEBD data, greater deviations arise at higher interaction strengths. For very large interaction strength (see for example $U=20\,\Delta$), the lengths of the central region $L$ considered here are not sufficient. At high bias voltage some spurious finite size effects of the reference system are visible in the form of peaks in the steady-state current.\\
\begin{figure}
	\centering	\includegraphics[width=0.49\textwidth]{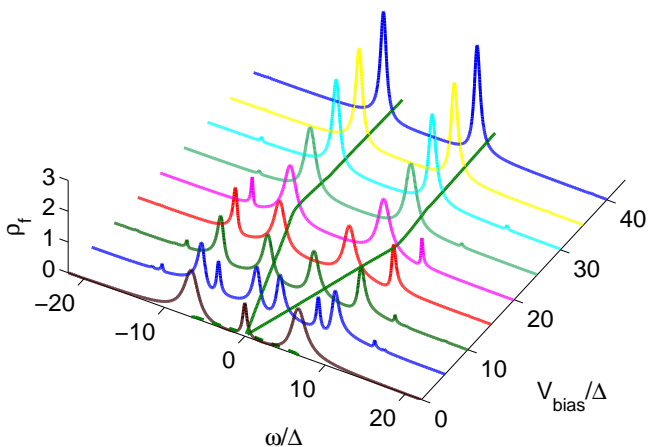}
	\caption{(Color online) Non-equilibrium local density of states of the interacting quantum dot. Data are shown for $U=12\,\Delta$ and an artificial numerical broadening of $0^+=0.05$. Results are obtained by nVCA$_{T}$ with $L=3$.}
	\label{fig:NDOSline}
\end{figure}
Next we would like to discuss the performance of nVCA. An illustration is shown in the mid column of \fig{fig:ssCurrent}. Here we compare nCPT and nVCA$_T$ for $L=3, 7 \mbox{ and }11$ again for increasing interaction strengths $U/\Delta=4,8,12 \mbox{ and }20$. We show a zoom to the low bias region where benchmarking data from other techniques, like TEBD data~\cite{unpubTEBD}, tDMRG~\cite{PhysRevB.79.235336}, FRG~\cite{jakobs_nonequilibrium_2010}, and QMC data obtained by Werner \etal~\cite{PhysRevB.81.035108} in the wide-band limit are available. One aspect to note immediately is that nVCA, like TEBD, departs sooner than nCPT from the linear response data, which is due to the better reproduction of the Kondo resonance within nVCA. The thinning of this narrow resonance at zero bias with increasing interaction strength is responsible for the departure from the linear response result~\cite{Dutt2011}. Since the Kondo resonance is better accounted for in nVCA, the curves leave the linear response data sooner than the respective nCPT results. The TEBD data~\cite{unpubTEBD} and QMC data obtained by Werner \etal~\cite{PhysRevB.81.035108} in the wide-band limit are also plotted and serve as a benchmark up to $V_{B}\approx5\,\Delta$ before effects of a different lead density of states become important. Curves obtained by tDMRG~\cite{PhysRevB.79.235336} / FRG~\cite{jakobs_nonequilibrium_2010} are not shown but lie basically on top of the TEBD / QMC data. Early data from perturbation theory~\cite{PhysRevB.68.155310} are known to have some additional spurious bumps in the low bias region. The improvement due to the variational feedback in nVCA becomes clear by comparing to the nCPT results for the same $L$, especially at higher interaction strengths.\\
\begin{figure}
	\centering
		\includegraphics[width=0.49\textwidth]{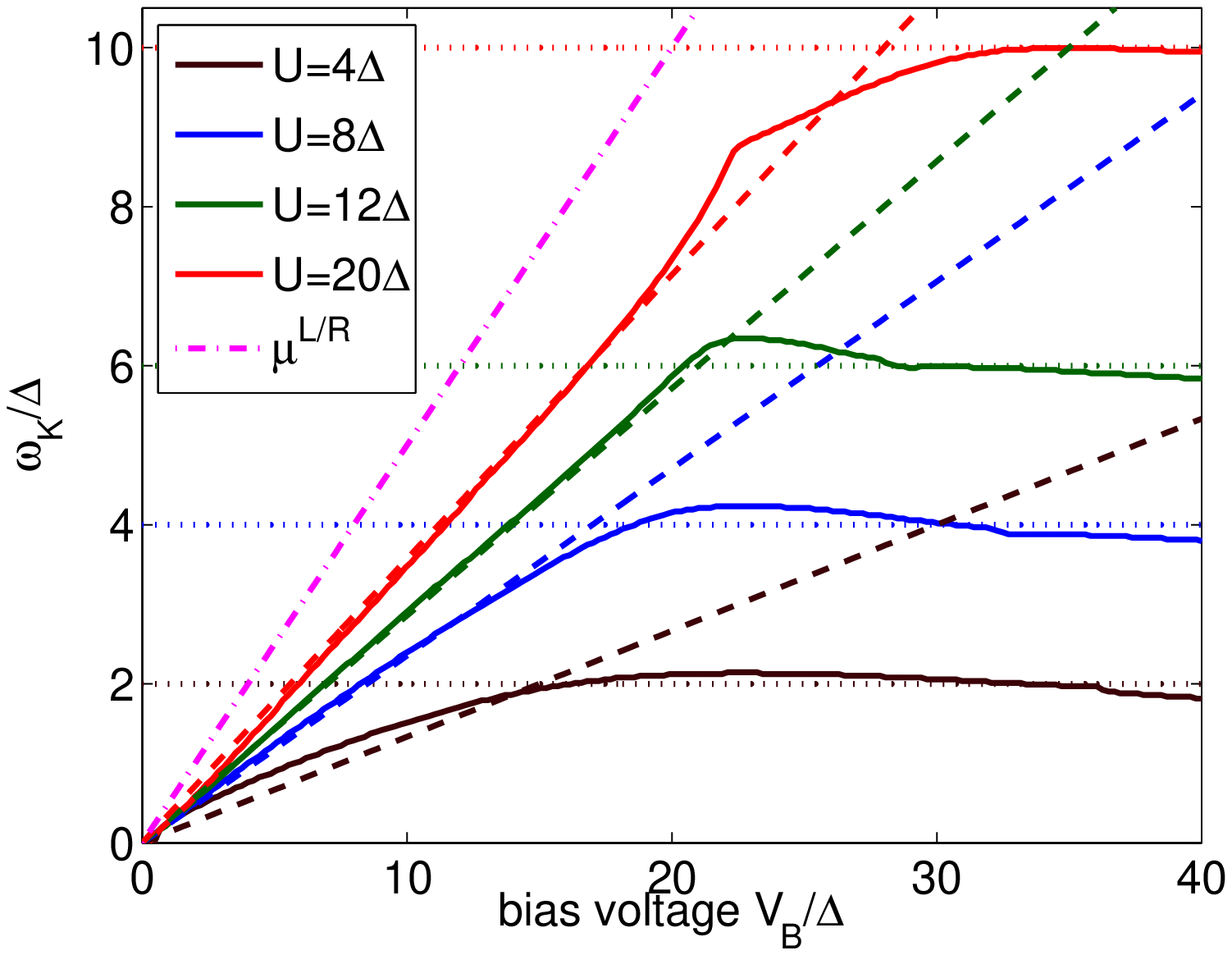}
	\caption{(Color online) Interaction dependent splitting of the Kondo resonance at $\omega_K$ under bias. Data shown are for different interaction strengths $U=4,8,12 \mbox{ and }20\,\Delta$ (solid lines). The resonance merges with the incoherent high energy spectrum (Hubbard bands), located at $\omega_H\approx \pm \frac{U}{2}$ (dotted lines) at a certain bias voltage $V_m\approx 15,17,21\mbox{ and }28$ respectively. These values have been obtained by the linear fits (indicated as dashed lines). If the Kondo resonance would pin at the chemical potential of the leads $\omega_K$, would follow the linear behavior indicated as $\mu_{L/R}$ (dash dotted line).}
	\label{fig:KondoSplit}
\end{figure}
The right column of \fig{fig:ssCurrent} shows a comparison of the performance of different variational parameters considered within nVCA for $L=7$. For small interaction strength all nVCA parameter sets work equally well. With increasing interaction strength, however, the different variational parameters predict a different behavior of the steady-state current. It should be noted here that some parameters like the ones used in nVCA$_{t_b}$ cause almost no deviation from the nCPT result while others improve it appreciably, like e.g. nVCA$_T$. It predicts a two peak structure in the high bias voltage regime, which however  is not observed directly in the TEBD benchmark data. No final conclusion can be drawn about the behavior of the current in this bias regime because TEBD can only predict upper bounds for the steady-state current there. The position of the dip in the steady-state current is at $V_B=U$ for all interaction strengths. We note that this is where the bare level position of the quantum dot $\epsilon_f=-\frac{U}{2}$ is about to stop overlapping with one of the lead density of states (while still overlapping with the other).\\
The calculation for two independent variational parameters (nVCA$_{t,t'}$) yields similar results as nVCA$_{T}$ in the lower bias region but goes to zero quickly for high bias voltages and does not show a dip.\\
On the whole, one may say that the self-consistent feedback implemented within nVCA provides a significant improvement over the nCPT results. This has been already observed in \tcite{PhysRevB.85.235107}. However, there this result was deduced from the convergence of results with increasing cluster size and not with a benchmark comparison with alternative numerical methods. For low interaction strengths $U\lesssim4\,\Delta$, bare nCPT already performs very well (independent of $L$) while for larger interaction strengths the variational improvement of nVCA becomes important. Motivated by the success of nVCA for the steady-state current, we proceed by evaluating the non-equilibrium local density of states of the quantum dot.\\
\begin{figure*}
	\centering
		\includegraphics[width=1.00\textwidth]{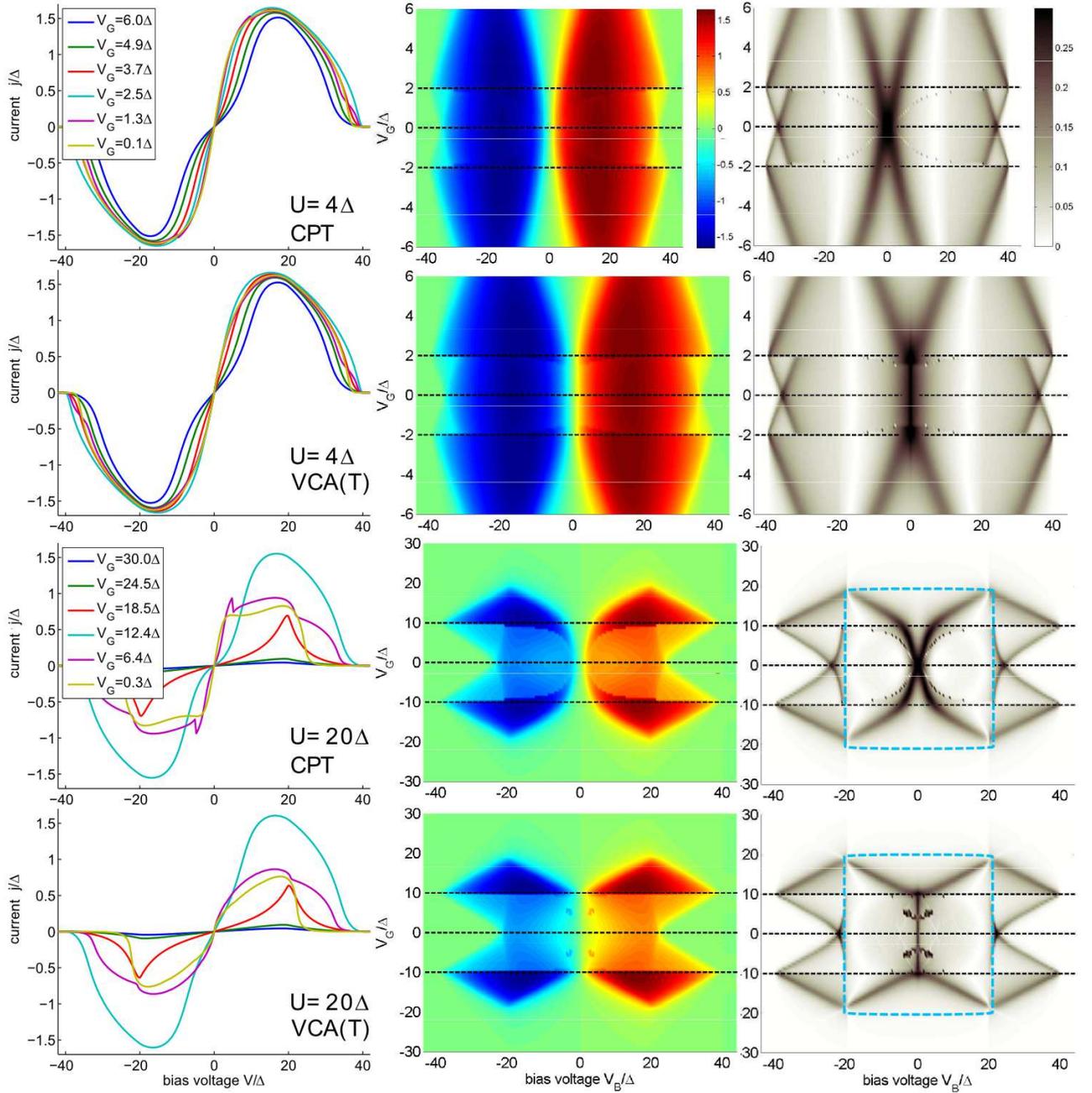}
	\caption{(Color online) Stability diagram and steady-state current as a function of bias voltage $V_{B}$ and gate voltage $V_G$. In the first column, the steady-state current is shown as a function of bias voltage for various gate voltages. The second column contains a density plot of the steady-state current in the full $V_{B} / V_{G}$ parameter space. In the third column a similar density plot is shown for the differential conductance $G=\frac{dI}{dV_{B}}$ (stability diagram). The four rows depict results obtained by nCPT for $U=4\,\Delta$, nVCA$_T$ for $U=4\,\Delta$, nCPT for $U=20\,\Delta$, and nVCA$_T$ for $U=20\,\Delta$. The squares in the two stability diagrams bottom right mark the experimentally most interesting region. In the nVCA$_T$ result, the Kondo region (vertical line in the center) is reproduced extremely well, as are the Coulomb blockade regions ($\propto V_B$) above and below. For comparison to recent experimental data see \tcite{PhysRevB.84.245316} (figure 5 therein). Effects at higher bias voltage arise from the finite bandwidth used here and are typically not be seen in experiment.}
	\label{fig:gatedCurrent}
\end{figure*}

\subsection{Non-equilibrium local density of states}\label{ssec:nLDOS}
The nLDOS in the quantum dot $\rho_f(\omega)$ as obtained by nCPT and nVCA$_T$ is depicted in \fig{fig:NDOS} for three sizes of the central part of the reference system: $L=3,7 \mbox{ and } 11$ and a large interaction strength of $U/\Delta = 12$ (corresponding to the steady-state current in \fig{fig:ssCurrent} (third row)). We plot the nLDOS in a density plot as a function of energy $\omega$ (horizontal) and applied bias voltage (vertical). The equilibrium result, consisting of a thin Kondo resonance at the Fermi energy ($\epsilon_F=0$ here) and two broad incoherent peaks located at $\approx \pm \frac{U}{2}$ with a width of $\approx2\,\Delta$, can be inferred from a horizontal cut at $V_{B}=0$. For finite bias voltage, a splitting of the Kondo resonance is observed in both nCPT (top row) and nVCA$_T$ (bottom row). It is well known that the noncrossing approximation (NCA) predicts a splitting of the Kondo resonance into two under voltage bias and that within second order perturbation theory the resonance is not split but suppressed only.~\cite{PhysRevB.68.155310} A linear splitting and slight broadening of the Kondo resonance with increasing bias voltage is proposed e.g. in \tcite{schiller_measuring_2001}. Intuitively it is expected that the split Kondo resonances pins at the chemical potentials of the leads. Several other methods yield a splitting with different features: real time diagrammatic~\cite{PhysRevB.54.16820} and scaling methods~\cite{PhysRevLett.90.076804} as well as the equation of motion technique~\cite{PhysRevLett.70.2601,PhysRevB.49.11040, PTPS.46.244, PTP.53.970, PTP.53.1286, PTP.54.316}. Within fourth order perturbation theory the Kondo resonance splits into two, which are located near the chemical potentials of the two leads~\cite{PhysRevB.68.155310}. In experiments on nano-devices a linear splitting of the Kondo resonance has been observed~\cite{PhysRevLett.89.156801, PhysRevLett.95.126603}. Such a linear splitting is also predicted by nVCA (see \fig{fig:NDOSline}), while nCPT yields a splitting which shows a roughly quadratic dependence on $V_B$. In addition within nVCA one observes an interaction dependent splitting (see \fig{fig:KondoSplit}). In contrast to the prediction of \tcite{schiller_measuring_2001} we do however not observe a simple pinning of the Kondo resonance at the chemical potentials of the leads. Our results indicate that the position $\omega_{\text{K}}$ of the split Kondo resonance depends on the interaction strength $U$: $\omega_{\text{K}}=\pm\frac{U}{2 V_m} V_{B} \text{ for } V_{B} \leq V_m$ and $\omega_{\text{K}}=\pm\frac{U}{2} \text{ for } V_{B} \geq V_m$ (see \fig{fig:KondoSplit}). Here $V_m$ is the voltage where the Kondo resonance merges with the high energy part of the spectrum located at $\omega_H\approx\pm\frac{U}{2}$: $V_m=15,17,21\mbox{ and } 28 \,\Delta$ for $U=4,8,12\mbox{ and } 20\,\Delta$. The $U$ dependent values of $V_m$ have been determined from the respective linear fit to the data.\\
For high voltages, these split peaks merge with the Hubbard bands and saturate, which has also been observed in fourth order perturbation theory calculations~\cite{PhysRevB.68.155310}. In this bias region a new low energy excitation is observed for $ L> 3$ within nVCA$_T$. This additional peak in the nLDOS has a dominant contribution to the total weight and is responsible for the two-peak structure observed in the nVCA$_T$ steady-state current (see \fig{fig:ssCurrent}).\\ 
\begin{figure}
	\centering	\includegraphics[width=0.49\textwidth]{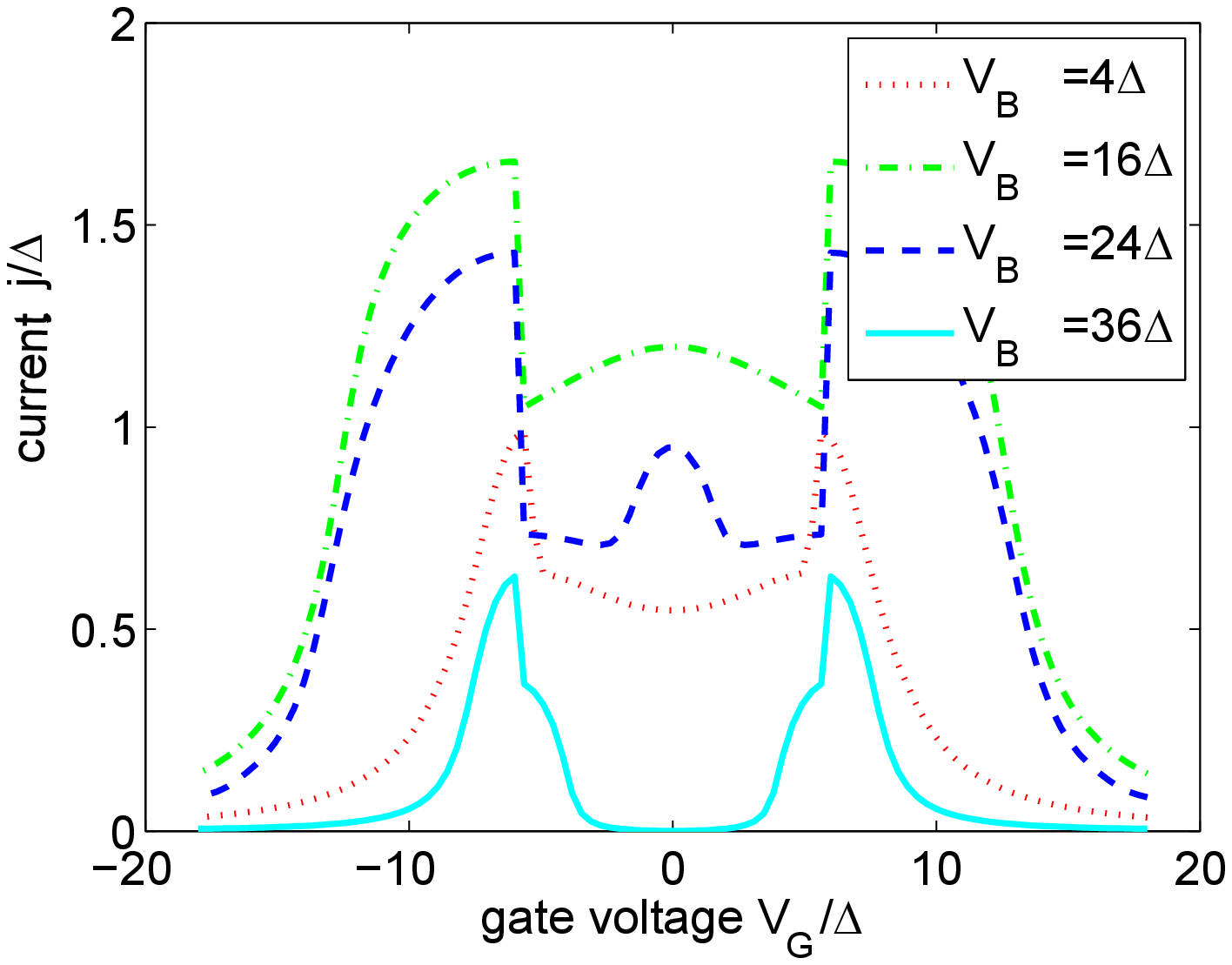}
	\caption{(Color online) Dependence of the steady-state current on gate voltage $V_G$. The largest steady-state current is obtained in a parameter regime which crosses from the Kondo plateau to the doubly occupied or empty  quantum dot. Data as obtained by nVCA$_T$ ($L=3$) are presented for an interaction strength of $U/\Delta=12$.}
	\label{fig:gatedCurrentofVg}
\end{figure}

\subsection{Finite gate voltage: steady-state current and stability diagram}\label{ssec:resultsGatedCurrent} 
It is interesting to investigate the steady-state current away from the particle-hole symmetric point and the region where Kondo correlations are present. In this section the current through the quantum dot under bias is analyzed as a function of an applied gate voltage $V_G$ and applied bias voltage $V_{B}$ at fixed interaction strength $U$. Results are presented for $L=3$. In this case no two-peak structure has been found in the nVCA$_T$ current in \fig{fig:ssCurrent}, which is corroborated by \fig{fig:NDOS} (bottom left). 
The dependence of the current on the gate- and bias voltage as obtained by nCPT and nVCA$_T$ is depicted in \fig{fig:gatedCurrent} for interaction strengths $U/\Delta=4$ and $U/\Delta=20$. First we discuss the so called stability diagram (differential conductance as a function of bias- and gate-voltage) which is shown in the third column of \fig{fig:gatedCurrent}. The dashed squares mark the region which is typically accessed experimentally. In the nVCA$_T$ result, the Kondo region (vertical line in the center) is reproduced extremely well, as are the Coulomb blockade regions ($\propto V_B$) above and below. Our data compare very well to recent experimental data of \tcite{PhysRevB.84.245316} (figure 5 therein). Effects at higher bias voltage arise from the finite bandwidth used here and are typically not seen in experiment.\\
The steady-state current as a function of bias and gate voltage is shown in the first and second column of \fig{fig:gatedCurrent}. It is interesting to observe that the largest current is obtained exactly at the crossover points from the Kondo to the empty or doubly occupied quantum dot (these regions are marked by black-dashed lines in middle and right panel of \fig{fig:gatedCurrent}). This aspect can be seen more clearly in \fig{fig:gatedCurrentofVg}, where the current is plotted  as a function of gate voltage. For $U=4\,\Delta$ there is not much difference between the nCPT and nVCA results as can be inferred from \fig{fig:gatedCurrent}. For $U=20\,\Delta$, however, we see that the feedback mechanism in VCA has a significant impact and leads to smoother $j-V_B$ curves due to suppression of finite size effects originating from the reference system. The sharp jumps at the crossover point in \fig{fig:gatedCurrentofVg} originate from abrupt changes of the particle number in the ground state of the central cluster $L=3$. We expect that this step smoothens with increasing $L$.\\
Concerning the reliability of the methods used, our data suggest that outside the parameter region where Kondo correlations dominate the single occupied quantum dot (in between the horizontal dashed black lines in \fig{fig:gatedCurrent}) nCPT and nVCA perform almost equally well. Results are significantly easier to obtain outside the Kondo plateau (which was mentioned before e.g. in \tcite{PhysRevB.85.235107} within exact diagonalization/CPT/VCA or in \tcite{heidrich_nonequilibrium_2010} within DMRG). Thereby the convergence within system size is greatly enhanced with respect to the Kondo region and very accurate results may be obtained already with small systems. Therefore, we may argue that for the steady-state of the SIAM, nCPT and nVCA perform quite well outside the Kondo region for any interaction strength as well as in the Kondo region for small interaction strengths. On the other hand, in the Kondo region nVCA outperforms nCPT for higher interaction strengths.\\

\section{Conclusions}\label{sec:conclusion}
We have presented results for the steady-state of the single-impurity Anderson model. We have applied non-equilibrium Cluster Perturbation Theory and its variational extension, the non-equilibrium Variational Cluster Approach to this model for a single quantum dot under bias. Both methods make use of the Keldysh Green's function formalism and are capable of working in the thermodynamic limit which is necessary to account for particle and energy dissipation mechanisms.\\
Results for the particle-hole symmetric model, which is dominated by Kondo correlations, have been compared to Time Evolving Block Decimation and quantum Monte Carlo. At low values of interaction strength they show excellent agreement already for non-equilibrium Cluster Perturbation Theory. For higher values of interaction strength, the self-consistency implemented within the non-equilibrium Variational Cluster Approach proves crucial in order to obtain reasonable results. Both methods coincide with the low bias linear response data for the steady-state current. Both methods furthermore become exact in the non-interacting limit.\\
The non-equilibrium local density of states of the quantum dot exhibits a linear and interaction dependent splitting of the bias voltage within the non-equilibrium Variational Cluster Approach which is not visible in the non-equilibrium Cluster Perturbation Theory. At a certain (interaction dependent) bias voltage we find that this split Kondo resonance merges with the high energy incoherent part of the spectrum.\\
When applying a gate voltage and thereby leaving the Kondo regime, calculations become a lot easier and non-equilibrium Cluster Perturbation Theory and the non-equilibrium Variational Cluster Approach appear to perform very well, which can be inferred from the convergence of our data. The highest current amplitude is obtained at the crossover from the Kondo to the empty or doubly-occupied quantum dot. Experimental stability diagrams are reproduced very well within the variational approach. They show a clear Kondo regime and a Coulomb blockade region.\\
We may conclude that the non-equilibrium Variational Cluster Approach is a promising method for the evaluation of steady-state quantities of strongly correlated model systems. Dynamic quantities become available in the whole complex plane and, in principle any fermionic or bosonic lattice model may be treated including multi-band or multi-orbital systems. It is both fast and versatile and in principle all parameter regimes of the model are accessible. Subjects for further investigation are the self-consistency criterion, which is not unique, as well as the use of different variational parameters and an optimized out of equilibrium variational principle.\\

\begin{acknowledgments}
We gratefully acknowledge fruitful discussions with S. Andergassen and thank P. Werner for providing his continuous time QMC data. This work was partly supported by the Austrian Science Fund (FWF) P24081-N16 and ViCoM sub projects F04103 and F04104. Some of the numerical calculations have been conducted at the Vienna Scientific Cluster (VSC-I$\&$II).\\
\end{acknowledgments}

\appendix
\section{Green's functions, notation and observables}\label{app:basics}
Upon introducing the basic notation we follow standard literature (see e.g. \tcite{rammer_quantum_1986}). The non-equilibrium single-particle properties are provided by the single-particle Green's function in the $2\times 2$ Keldysh space (denoted by tildes)
\begin{align}
  \widetilde{\GF} &= \begin{pmatrix} \GF^R & \GF^K \\ 0 & \GF^A \end{pmatrix}\;\mbox{,}
\label{eq:Gtilde}
\end{align}
where $\GF^{R/A/K}$ are again matrices in site/spin space and functions of two time coordinates~\cite{rammer_quantum_1986}. $R$ denotes the retarded, $A$ the advanced and $K$ the Keldysh component of the single-particle Green's function which for (non-relativistic) fermions read
\begin{align*}
  \GF^R_{ij}(\tau_1-\tau_2) &= \imath\theta(\tau_1-\tau_2)\langle[ c_i^\nag(\tau_1),c_j^\dag(\tau_2)]\rangle\\
  \GF^A_{ij}(\tau_1-\tau_2) &= -\imath\theta(\tau_2-\tau_1)\langle[ c_i^\nag(\tau_1),c_j^\dag(\tau_2)]\rangle\\
  \GF^K_{ij}(\tau_1-\tau_2) &= -\imath\langle c_i^\dag(\tau_1)c_j^\nag(\tau_2) + c_i^\nag(\tau_1)c_j^\dag(\tau_2)\rangle\,\mbox{,}
\end{align*}
where $i,j$ denote site as well as spin, $\tau_1,\tau_2$ real time and $[\hat{A},\hat{B}]=\hat{A}\hat{B}-\hat{B}\hat{A}$ the standard commutator. Note that in the following we consider the spin symmetric model (see \eq{eq:HSIAM}) and therefore suppress spin-indices. In the formulas for the current we therefore include an additional factor of two. The handy matrix relations $\GF^A(\tau_1-\tau_2)=(\GF^R)^{\dag}(\tau_2-\tau_1)$ and $\GF^K(\tau_1-\tau_2)=-(\GF^K)^{\dag}(\tau_2-\tau_1)$ follow from the definitions.\\
Our goal is to investigate the steady-state in which the system becomes time-translationally invariant. Therefore we may Fourier-transform to the energy domain and evaluate single-particle steady-state expectation values
\begin{align*}
  \langle c_i^{\dag} c_j\rangle &=\frac{\delta_{ij}}{2} + \frac{1}{2}\,\int_{-\infty}^\infty\,\frac{d\omega}{2\pi}\Im\text{m}\,\GF^K_{ij}(\omega)\mbox{.}
\end{align*}
The steady-state current through site $m$ can be evaluated~\cite{haug_quantum_1996, jauho_introduction_2006} by the time derivative of the total particle number to the left of this site 
\begin{align*}
 j &= \langle\dot{\hat{N}}_M(t)\rangle\;\mbox{,}\;\;\;\hat{N}_M=\sum\limits_{i=-\infty}^{m-1}\,\hat{n}_i\,\mbox{.}
\end{align*}
This expression for the current may be rewritten in terms of the Keldysh component of the single-particle Green's function
\begin{align}
 j_{i\, i+1} &= t_{i\, i+1}\,\int_{-\infty}^\infty\,\frac{d\omega}{2\pi}\,\Re\text{e} \left(\GF^K_{i\, i+1}(\omega) - \GF^K_{i+1\,i}(\omega)\right)\,\mbox{,} 
\label{eq:sscurrent}
\end{align}
where $t_{i\,i+1}$ is assumed to be real. The non-equilibrium density of states at site $i$ is given by 
\begin{align*}
 \rho_i(\omega) &= -\frac{1}{\pi}\,\Im\text{m}\,\GF_{ii}^R(\omega)\,\mbox{,}
\end{align*}
which has to fulfill the spectral sum rule
\begin{align}
 1 &\stackrel{!}{=} \int_{-\infty}^\infty\,\rho_i(\omega) \;\;\;\forall i\, \mbox{.}
\label{eq:ASR}
\end{align}

\section{Details on the nVCA procedure}\label{app:nvca}
In this paper we compare nCPT with nVCA for different auxiliary one particle terms $\hat{\Delta}(\matx)$. In agreement with our previous experience, that the variational hopping parameters are crucial for the physics in the Kondo regime~\cite{PhysRevB.85.235107}, we observe that it suffices to consider only hopping processes inside the central region of the reference system. The most general form of the auxiliary one-particle term considered in this work reads
\begin{align*}
\hat{\Delta}(\matx_{t},\matx_{t'},\matx_{t_{b}}) &= \matx_{t'} \sum_{\alpha,\sigma}
\big( c_{0\alpha\sigma}^{\dagger} \, f_{\sigma}^\nag +hc
\big)\\
&+\matx_{t}
\sum_{\alpha,\sigma}
\sum_{i=1}^{\ell-1} 
\, \left( c_{i\alpha\sigma}^{\dagger} \, c_{i-1\alpha\sigma}^{\nag} +hc\right)\\
&+\matx_{t_{b}}\sum_{\alpha,\sigma}  
\big( c_{\ell\alpha\sigma}^{\dagger} \, c_{\ell-1\alpha\sigma}^{\nag} 
+hc\big)\,\mbox{,}
\end{align*}
where $\ell\equiv (L-3)/2$.
It contains three different hopping processes. The first term
describes the hopping processes to and out of the quantum dot. The second term contains nearest neighbor hopping processes between the lead sites of the central region except the border one. The last hopping process involves the nearest neighbor sites at the boundary of the central region. In the sequel, we consider the following cases:
($\text{nVCA}_{T}$) with $\matx_{t'}=\matx_{t}=\matx_{t_{b}}$, 
($\text{nVCA}_{t,t'}$) with $\matx_{t_{b}}=0$,
($\text{nVCA}_{t'}$) with $\matx_{t} =\matx_{t_{b}}=0$, and
($\text{nVCA}_{t_{b}}$) with $\matx_{t} =\matx_{t'}=0$.\\
A bias voltage $V_{B}$ is applied in an anti-symmetric manner by setting the on-site energies as well as the chemical potentials of the left and right leads to $\epsilon_L = \mu_L = -\epsilon_R = -\mu_R = \frac{V_{B}}{2}$. Note that the sites of the leads which are incorporated in the central region of the reference system are also subjected to these on-site energies.
All calculations are done for sizes of the central region of $L=3, 7 \mbox{ and } 11$-sites,  which corresponds to symmetric central regions. We note that $L=5, 9$ suffer from a finite-size gap which closes with increasing $L$. This gap arises due to an even amount of sites to the left as well as to the right of the quantum dot.  We do not consider even $L$ since this would correspond to an unequal amount of sites of the left and right lead in the central region. Thereby, the geometric symmetry of the problem would be spoiled and the application of a bias voltage which respects particle-hole symmetry would not be possible.\\
In order to reduce finite size effects of the calculated quantities and to make an extrapolation to infinite central cluster sizes easier we scrutinized the idea of averaging over boundary conditions as outlined in \tcite{Gammel1992} or \tcite{LohJr1988A499}. 
In CPT/VCA one has the freedom to add single-particle terms to $\hat{h}$ and subtract them again in $\hat{\TF}$. Therefore we may add hopping terms   which connect the first and last site of the central cluster and which have an arbitrary phase. The spectral function of the impurity site for a central region with complex periodic boundary conditions is in general no longer particle hole symmetric, which makes it necessary to average $\widetilde{G}_{\phi}(\omega)$ over $\phi$ to retain the symmetry.\\
The result, however, was disappointing as the expected faster convergence towards the thermodynamic limit has not been confirmed.\\
In this work we do not consider the long-range part of the Coulomb interaction which leads to additional charging effects in real devices.\\

\end{document}